\documentclass[floatfix,aps,preprint,prl,pre,amssymb]{revtex4}
\usepackage{graphicx}
\usepackage{subfigure}

\begin{document}

\title{\textbf{The Darrieus-Landau instability \\ in fast deflagration and laser ablation}}

\author{Vitaly Bychkov, Mikhail Modestov, and Mattias Marklund}

\affiliation{Department of Physics, Ume\aa\ University, S--901 87
Ume\aa, Sweden}

\begin{abstract}
The problem of the Darrieus-Landau instability at a discontinuous
deflagration front in a compressible flow is solved. Numerous previous attempts to solve this problem
suffered from the deficit of boundary conditions.
Here, the required additional boundary condition is derived
rigorously taking into account the internal structure of the front.
The derived condition implies a constant mass flux at the front;
it reduces to the classical Darrieus-Landau condition in the limit
of an incompressible flow. It is demonstrated that in general the solution to the
problem depends on the type of energy source  in the flow. In the
common case of a strongly localized source, compression effects
make the Darrieus-Landau instability considerably weaker.
Particularly, the instability growth rate is
reduced for laser ablation in comparison with the classical
incompressible case. The instability disappears completely in the
Chapman-Jouguet regime of ultimately fast deflagration.
\end{abstract}


\maketitle

\section{I. Introduction}

The Darrieus-Landau (DL) instability is one of the most fundamental and important
instabilities in hydrodynamics \cite{Landau-Lifshitz-Fluid}. This
instability develops at a deflagration front, \textit{i.e.}, a front of
energy release propagating subsonically due to thermal conduction.
Among the many different examples of deflagrations we mention
chemical flames
\cite{Landau-Lifshitz-Fluid,Zeldovich.et.al-1985,Pelce-Clavin-1982,Searby-Clavin-1986,Bychkov-Liberman-2000,Kadowaki-Hasegawa-2005},
laser ablation in inertial confinement fusion and in manipulation of nanostructured surfaces
\cite{Bodner-1994,Bychkov.et.al-1994,Betti.et.al-1995,Betti.et.al-1996,Clavin-Masse-2004,Sanz.et.al-2006,Bychkov.et.al-2007,Garrison.et.al-2003,Plech.et.al-2006},
thermonuclear deflagration in supernovae
\cite{Woosley-Weaver-1986,Bychkov-Liberman-1995,Gamezo.et.al-2003,Bychkov.et.al-2006}, as well as
waves of phase transition. If we perturb an
initially planar deflagration front, the perturbations grow
because of the DL instability bending the front. At the nonlinear
stage, the DL instability increases the deflagration velocity;
recent reviews on the DL instability in flames may be found in, \textit{e.g.}, Refs.\
\cite{Bychkov-Liberman-2000,Kadowaki-Hasegawa-2005}. The classical
DL theory considers an infinitesimally thin deflagration front in
an incompressible flow \cite{Landau-Lifshitz-Fluid}. To the best
of our knowledge, the first attempt to study the influence of gas
compression on the DL instability was made in Ref.\
\cite{Kadowaki-1995}. Reference \cite{Kadowaki-1995} treated
compression effects as small corrections to the classical DL
theory. This is reasonable for chemical flames, since even
relatively fast laboratory flames have the Mach number much
smaller than unity both in the fuel mixture and in the burnt gas.
However, there is an important example of deflagration with
intrinsically strong compression effects; this is laser ablation
in inertial confined fusion
\cite{Manheimer.et.al-1982,Meyer-Thiell-1984,Fabbro.et.al-1985}.
 In laser
ablation, the isothermal Mach number is one at the critical
surface, where the laser radiation is partly absorbed and partly
reflected by the fusion plasma. Due to this, the DL instability in laser
ablation cannot be studied assuming small compression effects.

    In inertial confined fusion, the DL instability
typically appears together with the Rayleigh-Taylor (RT) instability
\cite{Bodner-1994,Bychkov.et.al-1994,Betti.et.al-1995,Betti.et.al-1996,Clavin-Masse-2004,Sanz.et.al-2006,Bychkov.et.al-2007}.
Quite often the DL instability is overwhelmed by the RT
instability because of the large acceleration of the plasma targets.
Still, taking sufficiently large targets, one obtains a small
acceleration, which tends to zero for an infinitely large target.
In that case the RT instability becomes weak, and one can observe
the DL instability of ablation flow in a pure form. The
DL instability in ablation flow was encountered in
 Ref.\ \cite{Bychkov.et.al-1994}
 within the model of incompressible flow (see Eq. (54) of the paper). However, Ref. \cite{Bychkov.et.al-1994} did not state this finding openly; for the first time it was done much later in Ref. \cite{Piriz-Portugues-2003}.
 Since then, there has been much interest to the DL instability in inertial
confined fusion both in the theory
\cite{He-2000,Clavin-Masse-2004,Sanz.et.al-2006,Keskinen.et.al-2007,Piriz-2001,Piriz-Portugues-2003}
and numerical simulations
\cite{Masse.et.al.-2000,Gotchev.et.al-2006,Goncharov.et.al-2006}.
 In the theoretical papers \cite{Bychkov.et.al-1994,Clavin-Masse-2004,Sanz.et.al-2006,Keskinen.et.al-2007}
the incompressible approximation was used, which
may be considered only as a qualitative model for the ablation flow.
The influence of compression effects on the DL instability in
laser ablation has been considered only recently \cite{Piriz-2001,Piriz-Portugues-2003}, where,
unfortunately, the plasma compression was treated only as
second-order corrections to the incompressible solution. Moreover,
the second-order corrections in
Refs.\ \cite{Piriz-2001,Piriz-Portugues-2003} concerned only terms
proportional to the deflagration thickness; this thickness is infinitesimal
in the classical DL theory. Such an approach can therefore not describe
properly the DL instability in laser ablation. One should expect
considerable modifications of the DL theory because of the flow
compression already for a discontinuous deflagration front. The
case of strong compression was considered in Ref. \cite{He-2000}.
However, Ref. \cite{He-2000} did not resolve properly the deficit
of boundary conditions at the deflagration front and
the solution was built on a hidden arbitrary assumption (see below). Thus, the
fundamental problem of the DL instability in fast deflagration
and laser ablation in a strongly compressible flow has remained without a solution so
far.

    The above problems with the DL theory in a
compressible flow are not surprising, cf.\ the RT instability in inertial confined
fusion. It is well-known that the model of a discontinuous
ablation front used to study the RT instability contains a deficit
in the boundary conditions
\cite{Bodner-1994,Bychkov.et.al-1994,Betti.et.al-1995,Betti.et.al-1996,Clavin-Masse-2004,Sanz.et.al-2006,Bychkov.et.al-2007}:
the number of unknown values exceeds the number of conservation
laws at the front by one. This deficit was obtained first in
Ref. \cite{Bodner-1994}; the way to overcome the trouble correctly was
suggested in Ref. \cite{Bychkov.et.al-1994} within the incompressible
model. In the case of incompressible flow the extra condition may
be easily guessed: it is the condition of a constant flame
velocity with respect to the fuel mixture
\cite{Landau-Lifshitz-Fluid}.
 This is the well-known DL condition;
it may be also proved rigorously for an isobaric flow, see \textit{e.g.}\ Refs.
\cite{Pelce-Clavin-1982,Searby-Clavin-1986,Bychkov-Liberman-2000}.
The problem of the additional boundary condition becomes difficult
again as we take compression effects into account.

    In order to overcome the trouble, Ref. \cite{Kadowaki-1995} assumed
an extra condition identical to the incompressible DL condition
(below we show that this was not the only assumption of
Ref. \cite{Kadowaki-1995}). As a result, Ref. \cite{Kadowaki-1995}
obtained that compression effects make the DL instability stronger
in the model of a discontinuous front. The papers
\cite{Piriz-2001,Piriz-Portugues-2003} used the same assumption as
Ref. \cite{Kadowaki-1995} in order to study the DL instability in an
ablation flow. Still, unlike Ref. \cite{Kadowaki-1995}, in
Refs. \cite{Piriz-2001,Piriz-Portugues-2003} no influence of the
compression effects for the discontinuous deflagration front was found.
Reference \cite{He-2000} demonstrated that the assumption of
Refs. \cite{Kadowaki-1995,Piriz-2001,Piriz-Portugues-2003} is incorrect:
it contradicts the conservation laws at the deflagration front. In
order to resolve the problem, Ref. \cite{He-2000} took into
account energy conservation at the front, which provided one extra
condition. Still, in that case one obtains also one extra
perturbation mode (the entropy mode), which means one extra
unknown value. Like before, the number of unknowns remains larger
than the number of equations by one. The solution proposed in
Ref. \cite{He-2000} reproduces the basic elements of reasoning of
Ref. \cite{Bodner-1994}. In order to overcome the deficit and to obtain
a solution to the spectral problem, Ref. \cite{He-2000} simply
omitted the entropy mode without any further explanation. This assumption
may be treated as a hidden additional boundary condition in the
analysis of Ref. \cite{He-2000}. Thus, we come to the question: what is
the correct additional boundary condition at a fast deflagration
front, taking into account a possibly strong compression of the plasma
or gas. Of course, this condition should go over to the DL condition in
the case of incompressible flow. When this question is answered,
we face the next issue: what is the influence of plasma compression
on the DL instability in laser ablation and fast deflagration? Does it increase the instability, as
suggested in Ref. \cite{Kadowaki-1995,Travnikov.et.al-1997,Travnikov.et.al-1999}, or leave it unchanged like
proposed in Refs. \cite{Piriz-2001,Piriz-Portugues-2003}? Or may be we
have the third option, and compression effects make the
instability weaker. The solution presented in Ref. \cite{He-2000} suggested a combination of
these tendencies. These questions are especially interesting for
the Chapman-Jouguet regime of fast deflagration, for
which the velocity of hot plasma or gas is equal to the local sound
speed, isothermal or adiabatic.

    In the present paper we solve the problem of the DL
instability at a discontinuous deflagration front in a
compressible flow. Taking into account the internal structure of the
front, we derive the additional boundary condition, which is
missing in the model of a discontinuous front. The derived
condition implies constant mass flux at the front;  it reduces
to the classical Darrieus-Landau condition in the limit of
incompressible flow. We demonstrate that the solution to the problem
in general depends on the type of energy source. In the
case of a strongly localized source, compression effects make the
DL instability considerably weaker. In particular, the DL
instability growth rate is reduced for laser ablation in
comparison with the classical incompressible case. The instability
disappears completely in the adiabatic Chapman-Jouguet regime of
ultimately fast deflagration.

\section{II. Basic equations for the DL instability}
\begin{figure}
{\includegraphics[width=1.1\columnwidth]{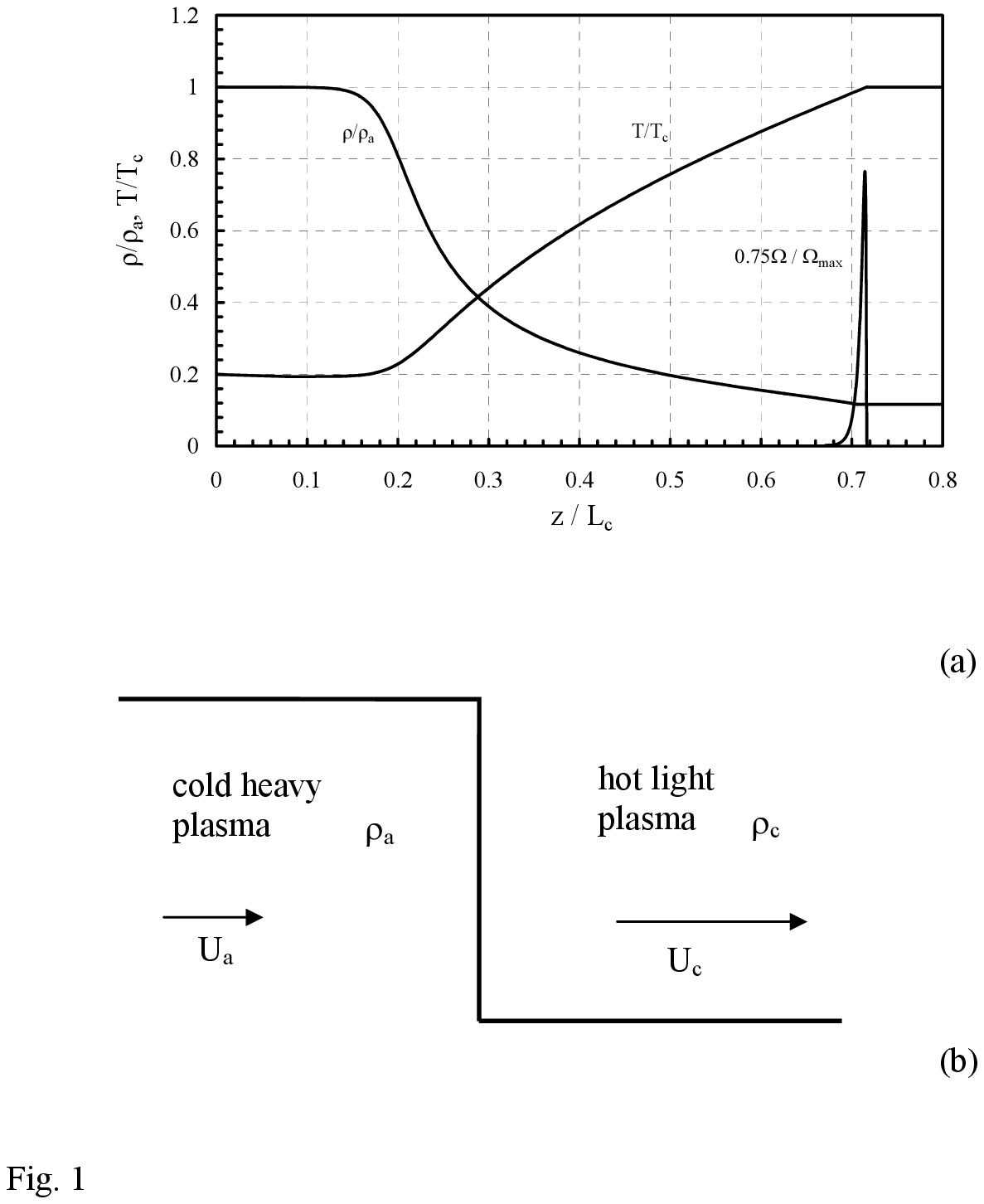}}
\caption{a)
Profiles of scaled density  $\rho / \rho_{a}$, temperature $T/T_{c}$
and energy release $0.75\Omega / \Omega_{max}$ for a deflagration
front with  $\Theta = 10$, $\kappa \propto T^{5/2}$, and
$Ma_{c}^{2}=1/\gamma$ (the isothermal CJ regime). b) The model of a
discontinuous deflagration front. } \label{fig-1}
\end{figure}

We start with basic hydrodynamic equations describing a plasma
(gas) flow in the absence of gravity
\begin{equation}
\label{eq1} \frac{\partial \rho }{\partial t} + \nabla\cdot (\rho
\textbf{u}) = 0,
\end{equation}
\begin{equation}
\label{eq2} \rho \frac{ \partial  \textbf{u}}{\partial t} + (\rho
\textbf{u} \cdot \nabla) \textbf{u} + \nabla P= 0,
\end{equation}
\begin{equation}
\label{eq3} \rho T \frac{ \partial  S}{\partial t} + \rho T
\textbf{u} \cdot \nabla S - \nabla \cdot (\kappa \nabla T)=
\Omega_{R},
\end{equation}
and the equation of state of an ideal gas
\begin{equation}
\label{eq4} P = \frac{\gamma - 1}{\gamma} C_{P}\rho T .
\end{equation}
We take the equation of energy transfer in the form describing
variations of entropy
\begin{equation}
\label{eq5} S = C_{V} \ln \left(\frac{P}{\rho^{\gamma}}\right) ,
\end{equation}
 where  $C_{P}$, $C_{V}$   are heat capacities at constant volume and
 pressure, $\gamma = C_{P}/C_{V}$
is the adiabatic exponent and  $\kappa$ is the coefficient of
thermal conduction. In general, kinetics of energy release in a
deflagration flow may involve additional differential equations
specific for a particular type of deflagration. Since it is
impossible to consider all particular cases in one paper, here we
take energy release described by some function  $\Omega_{R} (\rho,
T)$. Most of the results of the present work do not depend on
$\Omega_{R} (\rho, T)$; we only demand that energy release is
strongly localized in a narrow zone inside the deflagration front,
which is the usual case. This demand concerns not only the
function  $\Omega_{R}$, but also the derivatives
$\partial\Omega_{R}/ \partial\rho$,  $\partial\Omega_{R}/\partial
T$. In the case of laser ablation the energy release is typically
presented by $\delta$-function, which implies sufficiently strong
localization, e.g. see Refs. \cite{Bodner-1994,Bychkov.et.al-1994,Manheimer.et.al-1982}. We also assume that the function  $\Omega_{R}$
allows a planar stationary solution consisting of two uniform
flows of cold heavy plasma (label "a") and hot light plasma (label
"c") separated by a transitional region, which is the deflagration
front. The labels "a" and "c" originate from the ablation and
critical surfaces in the laser deflagration. Typical internal
structure of the deflagration front is illustrated in Fig.
\ref{fig-1}. Strictly speaking, laser ablation requires the flow
of hot light plasma in the form of a rarefaction wave, not a
uniform flow \cite{Manheimer.et.al-1982}. Still, by assuming a
uniform flow, we may consider all possible values of the Mach
number in the light plasma from zero to unity, which would be
impossible with a rarefaction wave. Besides, the model of two
uniform flows separated by the transition region is quite common
in the studies of the DL and RT instabilities in laser ablation,
e.g. see Refs.
\cite{Bodner-1994,Bychkov.et.al-1994,Piriz-2001,Piriz-Portugues-2003,He-2000}.

    The planar stationary deflagration may be described by the integrals
\begin{equation}
\label{eq6} \rho u_{z}=\rho_{a} U_{a}=\rho_{c} U_{c},
\end{equation}
\begin{equation}
\label{eq7} P+ \rho u_{z}^{2}=P_{a}+\rho_{a}
U_{a}^{2}=P_{c}+\rho_{c} U_{c}^{2}.
\end{equation}
One of the main dimensionless parameters in the problem is the
expansion factor
\begin{equation}
\label{eq8} \Theta= \frac{
\rho_{a}}{\rho_{c}}=\frac{U_{c}}{U_{a}}.
\end{equation}
In the case of ablation flow, the laser frequency determines the
critical density and the expansion factor. The other important
parameter is the Mach number in the light plasma (gas)
corresponding to the adiabatic sound
\begin{equation}
\label{eq9} Ma_{c}^{2}= \frac{ \rho_{c}U_{c}^{2}}{\gamma P_{c}}.
\end{equation}
In laser ablation, the isothermal Mach number is equal unity
$\rho_{c}U_{c}^{2}/P_{c}=1$ in the light plasma, and we have the adiabatic Mach number
$Ma_{c}^{2} = 1/\gamma$. Still, in the
present work we consider a general case of an arbitrary Mach
number between zero and unity. The Mach number in the heavy plasma
follows from (\ref{eq7}) as
\begin{equation}
\label{eq10} Ma_{a}^{2}= \frac{ Ma_{c}^{2}}{\Theta + \gamma
(\Theta - 1) Ma_{c}^{2}}.
\end{equation}
The internal structure of the deflagration front obeys the
stationary equation of energy transfer
\begin{equation}
\label{eq11} \rho_{a} U_{a} T \frac{dS}{dz} -
\frac{d}{dz}\left(\kappa \frac{dT}{dz} \right)= \Omega_{R}.
\end{equation}
Characteristic width of the deflagration front is determined by
thermal conduction in the hot plasma with typical definition
\begin{equation}
\label{eq12} L_{c}\equiv   \frac{\kappa_{c}}{C_{P}\rho_{a} U_{a}}=
\frac{\kappa_{c}}{C_{P}\rho_{c} U_{c}} .
\end{equation}

Small perturbations $\tilde{\varphi}$  to the stationary solution
$\varphi (z)$ have the general form
\begin{equation}
\label{eq13} \tilde{\varphi} (x,z,t) = \tilde{\varphi} (z)
\exp(\sigma t + ikx) ,
\end{equation}
where  $k$ is the perturbation wave number and  $\sigma$ is the
instability growth rate. In the problem of the DL instability
$\sigma$  is a real positive value. Then linearized system
(\ref{eq1})  - (\ref{eq4}) is
\begin{equation}
\label{eq14} \sigma \tilde{\rho}  + \frac{d}{dz} (\tilde{\rho}
u_{z} + \rho \tilde{u}_{z}) + ik \rho \tilde{u}_{x} = 0 ,
\end{equation}
\begin{equation}
\label{eq15} \sigma \rho \tilde{u}_{x}  + \rho u_{z}
\frac{d\tilde{u}_{x}}{dz} + ik \tilde{P} = 0 ,
\end{equation}
\begin{equation}
\label{eq16} \sigma \rho \tilde{u}_{z}  + \rho u_{z}
\frac{d\tilde{u}_{z}}{dz} + (\tilde{\rho} u_{z} + \rho
\tilde{u}_{z}) \frac{d u_{z}}{dz} + \frac{d\tilde{P}}{dz} = 0,
\end{equation}
\begin{eqnarray}
&&\sigma \rho T \tilde {S} + (\tilde{\rho} u_{z} + \rho
\tilde{u}_{z}) T \frac{d S}{dz} + \rho_{a} U_{a} \tilde{T}
\frac{dS}{dz} + \rho_{a} U_{a} T \frac{d \tilde{S}}{dz}
\nonumber \\
&& - \frac{d^{2}}{dz^{2}}(\kappa \tilde{T}) + k^{2}\kappa
\tilde{T} = \frac{\partial \Omega_{R}}{\partial \rho }
\tilde{\rho} + \frac{\partial \Omega_{R}}{\partial T } \tilde{T},
 \label{eq17}
\end{eqnarray}
\begin{equation}
\label{eq18} \tilde{P} = \frac{\gamma - 1}{\gamma} C_{P}
(\tilde{\rho} T + \rho \tilde{T}),
\end{equation}
\begin{equation}
\label{eq18a} \tilde{S} = C_{V} \left(\frac{\tilde{P}}{P} -
\gamma\frac{\tilde{\rho}}{\rho} \right).
\end{equation}
Since the instability develops at the front, then solution to
(\ref{eq14}) - (\ref{eq18a}) should decay at infinity ($z
\rightarrow \pm \infty$) in the uniform flows of heavy and light
plasma. All coefficients of Eqs. (\ref{eq14}) - (\ref{eq18a}) are
constant in the uniform flows, where the perturbations take the form
\begin{equation}
\label{eq19} \tilde{\varphi} (x,z,t) = \tilde{\varphi} \exp(\sigma
t + ikx + \mu z) ,
\end{equation}
with  $\mu > 0$ in the heavy plasma  $z \rightarrow - \infty$ and
$\mu < 0$ in the light plasma  $z \rightarrow  \infty$. Then, in
the uniform flows, the system (\ref{eq14}) - (\ref{eq18a}) reduces
to
\begin{equation}
\label{eq20} \sigma \tilde{\rho}  + \mu (\tilde{\rho} u_{z} + \rho
\tilde{u}_{z}) + ik \rho \tilde{u}_{x} = 0 ,
\end{equation}
\begin{equation}
\label{eq21} \sigma \rho \tilde{u}_{x}  + \rho u_{z} \mu
\tilde{{u}}_{x}
 + ik \tilde{P} = 0 ,
\end{equation}
\begin{equation}
\label{eq22} \sigma \rho \tilde{u}_{z}  + \rho u_{z} \mu
\tilde{u}_{z}  + \mu \tilde{P} = 0,
\end{equation}
\begin{eqnarray}
\sigma \rho T \tilde {S}   + \rho_{a} U_{a} T \mu \tilde{S} -
\mu^{2}\kappa \tilde{T}+ k^{2}\kappa \tilde{T} = \frac{\partial
\Omega_{R}}{\partial \rho } \tilde{\rho} + \frac{\partial
\Omega_{R}}{\partial T } \tilde{T}.
 \label{eq23}
\end{eqnarray}
In the case of a strongly localized energy source we have
$\partial\Omega_{R}/ \partial\rho = 0$,
$\partial\Omega_{R}/\partial T = 0$   in the uniform flows, and
the right-hand side in Eq. (\ref{eq23}) is simply zero. In the
present work we are interested mainly in energy sources of this type. Still,
another situation is also possible. For example, Refs.
\cite{Travnikov.et.al-1997,Travnikov.et.al-1999} considered a
first-order reaction of the Arrhenius type. In that case
perturbations of the energy source are not zero in the hot gas;
they may influence structure of acoustic modes and modify the whole
solution to the problem. As we show below, such modifications take
place even in the model of a discontinuous deflagration front.
Then, in order to solve the problem, one has to specify particular
kinetics of energy release, which is beyond the scope of the
present work.

\section{III. The model of a discontinuous deflagration front}

\textbf{\it{A. Solution in the uniform flows}}

In the present work we solve the stability problem within the
classical DL model of a discontinuous deflagration front,  $k
L_{c} << 1$. The traditional scaling for the DL instability growth
rate is
\begin{equation}
\label{eq24} \sigma = \Gamma U_{a} k  .
\end{equation}
Our purpose is to find the coefficient $\Gamma$  as a function of
the expansion factor $\Theta$  and the Mach number in the hot
plasma  $Ma_{c}$. Within the discontinuity model we have to solve
only Eqs. (\ref{eq14}) - (\ref{eq16}). The equation of energy
transfer Eq. (\ref{eq17}) is out of the model, since it contains
thermal conduction and energy release "hidden" inside the
infinitesimally thin front. Still, as we will see below, the
discontinuity model involves uncertainties, which make the
solution ambiguous and which may be removed correctly only by
taking into account Eq. (\ref{eq17}). First of the uncertainties
is the relation between pressure and density perturbations in the
so-called "sound" modes in the uniform flows. Previous works on
the subject Refs.
\cite{Kadowaki-1995,Piriz-2001,Piriz-Portugues-2003,He-2000}
assumed the adiabatic relation
\begin{equation}
\label{eq25} \tilde{P} = \gamma \frac{P}{\rho} \tilde{\rho} =
c_{s}^{2}\tilde{\rho}.
\end{equation}
As we can find from Eq. (\ref{eq23}), this is indeed the case for
a strongly localized energy source with  $\partial\Omega_{R}/
\partial\rho = 0$, $\partial\Omega_{R}/\partial T = 0$    in the
uniform flows. The power exponent  $\mu$ in Eqs. (\ref{eq19}) -
(\ref{eq23}) scales for the sound modes as $\mu \propto k$, see
below. Within the limit of discontinuous deflagration front, the
terms with heat conduction in Eq. (\ref{eq23}) become as small as
$k L_{c} << 1$, and the equation describes drift of entropy
perturbations:
\begin{equation}
\label{eq26} \rho T (\sigma + u_{z} \mu)\tilde{S} = 0.
\end{equation}
The combination $(\sigma + u_{z} \mu)$  stands for drift with the
flow, and it is non-zero for the sound modes (see the calculations
below). Then Eq. (\ref{eq26}) leads to zero perturbations of
entropy in the sound modes in agreement with Eq. (\ref{eq25}).
We stress that this is not a general case, but it holds
only for a strongly localized energy source. In the opposite case
of non-zero derivatives  $\partial\Omega_{R}/
\partial\rho $, $\partial\Omega_{R}/\partial T $ in (\ref{eq23}), the perturbations of energy
release become dominating in (\ref{eq23}) in the uniform flow of
hot plasma. In that case the relation of $\tilde{P}$  and
$\tilde{\rho}$ is determined by a particular structure of the
energy source. We repeat one more time that in the present work we
are interested mainly in strongly localized sources with
$\partial\Omega_{R}/
\partial\rho $ = 0, $\partial\Omega_{R}/\partial T = 0$   in the uniform
flows, which leads to the adiabatic relation (25) in the sound
modes.

    With these results in mind, we can obtain an equation for the power
    exponent $\mu$  from (\ref{eq21}) - (\ref{eq23})
\begin{equation}
\label{eq27} (\sigma + u_{z} \mu) \left[ (\sigma + u_{z} \mu)^{2}
+ c_{s}^{2} (k^{2}- \mu^{2})\right]=0 .
\end{equation}
 Solving (\ref{eq27}), we find possible perturbation modes in the uniform
flows describing vorticity drift
\begin{equation}
\label{eq28} \mu_{V}=-\frac{\sigma}{U_{c}} < 0 ,
\end{equation}
and sound
\begin{equation}
\label{eq29} \mu_{a,c}=-\frac{\sigma}{U_{a,c}}
\frac{Ma^{2}}{1-Ma^{2}}\pm
\sqrt{\frac{\sigma^{2}}{U_{a,c}^{2}}
\frac{Ma^{2}}{(1-Ma^{2})^{2}}+\frac{k^{2}}{1-Ma^{2}}}
,
\end{equation}
with $\mu_{a}> 0$  and  $\mu_{c}< 0$  for "$+$" and "$-$" in
(\ref{eq29}) corresponding to perturbations in the cold heavy and
hot light plasma, respectively. Mark that Eq. (\ref{eq29}) is different
from the respective results in Ref. \cite{Kadowaki-1995,Piriz-2001},
which were written as expansion of (\ref{eq29}) in powers of small
Mach number,  $Ma<<1$. Still, the limit of $Ma<<1$  does not hold
for the ablation flow. The structure of the perturbation modes in the uniform plasma follows from
Eqs. (\ref{eq21}) - (\ref{eq23}) as:
\begin{equation}
\label{eq30} \tilde{P}_{a} = - \rho_{a} U_{a}
\left(\frac{\sigma}{U_{a}\mu_{a} }+ 1\right) \tilde{u}_{za},
\end{equation}
\begin{equation}
\label{eq31} \tilde{u}_{xa} = i\frac{k}{\mu_{a}} \tilde{u}_{za},
\end{equation}
 for the sound mode in the cold plasma;
\begin{equation}
\label{eq32} \tilde{P}_{s} = - \rho_{c} U_{c}
\left(\frac{\sigma}{U_{c}\mu_{c} }+ 1\right) \tilde{u}_{zs},
\end{equation}
\begin{equation}
\label{eq33} \tilde{u}_{xs} = i\frac{k}{\mu_{c}} \tilde{u}_{zs},
\end{equation}
for the sound mode in the hot plasma; and
\begin{equation}
\label{eq34} \tilde{P}_{V} = 0,
\end{equation}
\begin{equation}
\label{eq35} \tilde{u}_{xV} = i\frac{\sigma}{U_{c}k}
\tilde{u}_{zV},
\end{equation}
for the vorticity mode in the hot plasma. Density perturbations
are related to pressure perturbations by Eq. (\ref{eq25}).
Equations (\ref{eq32}) - (\ref{eq35}) in the hot plasma may be
also reduced to one condition
\begin{equation}
\label{eq36} i \tilde{u}_{xc}
=\left(\frac{k}{\mu_{c}}+\frac{\sigma}{ U_{c}k}\right)
\left(\frac{\sigma}{U_{c}\mu_{c}}+1\right)^{-1}
 \frac{\tilde{P}_{c}}{\rho_{c} U_{c}}
+\frac{\sigma}{U_{c}k } \tilde{u}_{zc}
\end{equation}
 for  $\tilde{u}_{zc}=\tilde{u}_{zs}+\tilde{u}_{zV}$,
 $\tilde{u}_{xc}=\tilde{u}_{xs}+\tilde{u}_{xV}$,
 $\tilde{P}_{c}=\tilde{P}_{s}$.

\textbf{\it{B. Conditions at the discontinuous deflagration
front}}

The solution in the uniform flows (\ref{eq30}) - (\ref{eq35}) has
to be matched at the perturbed deflagration front $z_{f}= f
\exp(ikx + \sigma t)$ using the conservation laws of mass and
momentum \cite{Landau-Lifshitz-Fluid}
\begin{equation}
\label{eq37} \tilde{\rho}_{a}U_{a} +
\rho_{a}\left(\tilde{u}_{za}-\frac{\partial f}{\partial t}\right)=
\tilde{\rho}_{c}U_{c} +
\rho_{c}\left(\tilde{u}_{zc}-\frac{\partial f}{\partial t}\right),
\end{equation}
\begin{equation}
\label{eq38} \tilde{u}_{xa}+U_{a}\frac{\partial f}{\partial x} =
\tilde{u}_{xc}+U_{c}\frac{\partial f}{\partial x},
\end{equation}
\begin{equation}
\label{eq39} \tilde{P}_{a}+\tilde{\rho}_{a}U_{a}^{2} +
2\rho_{a}U_{a}\left(\tilde{u}_{za}-\frac{\partial f}{\partial
t}\right)= \tilde{P}_{c}+\tilde{\rho}_{c}U_{c}^{2} +
2\rho_{c}U_{c}\left(\tilde{u}_{zc}-\frac{\partial f}{\partial
t}\right).
\end{equation}
One can find similar equations in Refs. \cite{Kadowaki-1995,He-2000}. At
this point one can check that Eqs. (\ref{eq37}) - (\ref{eq39})
 are not sufficient to solve the problem. Indeed, we
have only 3 equations for 4 unknown values: the mode amplitudes
$\tilde{u}_{za}$, $\tilde{u}_{zs}$, $\tilde{u}_{zV}$, and the
front perturbation  $f$. Here we face the deficit of boundary
conditions at the ablation front known in the problem of the RT
instability in inertial confined fusion
\cite{Bodner-1994,Bychkov.et.al-1994}: in order to solve the
problem we need one more condition at the front. Strictly
speaking, one faces the same problem in the flame stability
theory. Still, in the classical DL case of incompressible flow the
missing condition may be easily guessed: this is the condition of
a constant velocity of flame propagation with respect to the fuel
mixture
\begin{equation}
\label{eq40} \tilde{u}_{za}-\frac{\partial f}{\partial t}=0.
\end{equation}
Ref. \cite{Kadowaki-1995} assumed a condition identical to
(\ref{eq40}) for a flame in a compressible flow. Surprisingly,
that was not the only assumption of paper \cite{Kadowaki-1995}. In
addition to (\ref{eq40}), Ref. \cite{Kadowaki-1995} assumed also a
similar condition for the hot gas
\begin{equation}
\label{eq41} \tilde{u}_{zc}-\frac{\partial f}{\partial t}=0.
\end{equation}
In the limit of incompressible flow Eq. (\ref{eq37}) reduces to
\begin{equation}
\label{eq42} \rho_{a}\left(\tilde{u}_{za}-\frac{\partial
f}{\partial t}\right)= \rho_{c}\left(\tilde{u}_{zc}-\frac{\partial
f}{\partial t}\right),
\end{equation}
so that Eq. (\ref{eq41}) follows from (\ref{eq40}).  However, in
the case of compressible flow, these two equations are different.
As a result, paper \cite{Kadowaki-1995} suggested 5 equations for
4 unknowns, which made the problem over-defined. The paper
\cite{He-2000} criticized the solution \cite{Kadowaki-1995}, but
it did not avoid an arbitrary assumption either. The paper
\cite{He-2000} complemented the system (\ref{eq37}) - (\ref{eq39})
by the perturbed equation of energy conservation, and the system
of modes (\ref{eq30}) - (\ref{eq35}) by the entropy mode
(\ref{eq26}). As a result, the approach of Ref. \cite{He-2000} involved
4 equations and 5 unknowns. In order to obtain the solution, one
mode was neglected in Ref. \cite{He-2000}; namely, the entropy mode.
That was just another arbitrary assumption; no physical
explanation was provided in Ref. \cite{He-2000} for this step. Thus, we
come to the problem of correct additional boundary condition at a
discontinuous deflagration front, which should replace
(\ref{eq40}) in the case of compressible flow.

    The missing equation may be derived only taking
into account the perturbed equation of energy transfer, Eq.
(\ref{eq17}). The term  $k^{2}\kappa \tilde{T}$ in Eq.
(\ref{eq17}) may be neglected since it is small as $kL_{c}<<1$  in
the approximation of a discontinuous front. We split other terms
into two groups:
\begin{eqnarray}
\rho_{a} U_{a} \tilde{T} \frac{dS}{dz} + \rho_{a} U_{a} T \frac{d
\tilde{S}}{dz} - \frac{d^{2}}{dz^{2}}(\kappa \tilde{T}) -
\frac{\partial \Omega_{R}}{\partial \rho } \tilde{\rho} -
\frac{\partial \Omega_{R}}{\partial T } \tilde{T},
 \label{eq43}
\end{eqnarray}
and
\begin{eqnarray}
\sigma \rho T \tilde {S} + (\tilde{\rho} u_{z} + \rho
\tilde{u}_{z}) T \frac{d S}{dz} .
 \label{eq44}
\end{eqnarray}
We notice that perturbations in the form of front bending
\begin{equation}
\label{eq45} \tilde{T}= -f \frac{d T}{d z},
 \quad \tilde{\rho}= -f \frac{d \rho}{d z},
\quad \tilde{S}= -f \frac{d S}{d z}
\end{equation}
turn the group Eq. (\ref{eq43}) to zero. To find this, we take
z-derivative of Eq. (\ref{eq11}) for the stationary deflagration
front. We also notice that perturbations of density and
temperature in the sound modes (\ref{eq30}) - (\ref{eq33}) make
the next order corrections in $kL_{c}<<1$  to the solution
(\ref{eq45}). We substitute (\ref{eq45}) into the perturbed
continuity equation Eq. (\ref{eq14}), integrate (\ref{eq14}) over
the deflagration front for zero-order terms in $kL_{c}<<1$ , and
find
\begin{equation}
\label{eq46} (\tilde{\rho} u_{z} + \rho \tilde{u}_{z}) -
(\tilde{\rho} u_{z} + \rho \tilde{u}_{z})_{a}= \sigma f (\rho -
\rho_{a}),
\end{equation}
which corresponds to the perturbed equation of mass conservation
at the front, Eq. (\ref{eq37}). In a similar way, substituting Eq.
(\ref{eq45}) into (\ref{eq15}), (\ref{eq16}) and integrating, we
may obtain the perturbed conservation laws (\ref{eq38}),
(\ref{eq39}). As a next step we have to check, if the solution
(\ref{eq45}), (\ref{eq46}) turn the second group, Eq.
(\ref{eq44}), to zero. Substituting the solution (\ref{eq45}),
(\ref{eq46}) into (\ref{eq44}) we find
\begin{eqnarray}
\sigma \rho T \tilde {S} + (\tilde{\rho} u_{z} + \rho
\tilde{u}_{z}) T \frac{d S}{dz}=\left[(\tilde{\rho} u_{z} + \rho \tilde{u}_{z})_{a}- \sigma f
\rho_{a} \right] T \frac{d S}{dz} .
 \label{eq47}
\end{eqnarray}
This combination turns to zero under the condition
\begin{eqnarray}
(\tilde{\rho} u_{z} + \rho \tilde{u}_{z})_{a}- \sigma f \rho_{a}
=0.
 \label{eq48}
\end{eqnarray}
Thus, solution in the form of front bending (\ref{eq45}) satisfies
the equation of energy transfer with the additional condition
\begin{equation}
\label{eq49} \tilde{\rho}_{a}U_{a} +
\rho_{a}\left(\tilde{u}_{za}-\frac{\partial f}{\partial
t}\right)=0.
\end{equation}
Equation (\ref{eq49}) plays the role of the DL condition modified
for a compressible flow. In the incompressible limit,  $Ma <<1$,
density perturbations are negligible and Eq. (\ref{eq49}) goes
over to the classical DL condition Eq. (\ref{eq40}). The physical
meaning of Eq. (\ref{eq49}) is zero perturbation of the mass flux,
which is mass burning rate in combustion or ablation rate in laser
fusion. It is interesting that the obtained additional condition,
Eq. (\ref{eq49}), does not depend on the type of energy release,
see the derivation. We also stress that Eq. (\ref{eq49}) is not a
conservation law, and it should not be confused with energy
conservation, though it was obtained from the equation of energy
transfer. On the contrary, Eq. (\ref{eq49}) follows from the
equation of energy transfer as an eigenvalue, similar to the
incompressible case \cite{Bychkov-Liberman-2000}, or even to the
planar stationary Zeldovich - Frank-Kamenetski solution
\cite{Zeldovich.et.al-1985}. This result should be expected, since
the deflagration speed (mass burning rate) is an eigenvalue of the
basic equations (\ref{eq1}) - (\ref{eq3}), which does not follow
from the conservation laws \cite{Landau-Lifshitz-Fluid}. By this
reason, the deflagration speed has to be obtained as an eigenvalue both for the
unperturbed solution and for perturbations. This problem is
intrinsic for deflagration, which is a discontinuity separating
two subsonic flows. The problem of additional conditions does not
arise in the case of detonation, for which one of the flows is
supersonic. Unlike deflagration, dynamics of a detonation front is
determined completely by the conservation laws. Thus, the
reasoning of Ref. \cite{He-2000} should work for detonation, but not
for deflagration. The condition Eq. (\ref{eq49}) leads to an
interesting consequence. Substituting (\ref{eq49}) into the
equation of dynamical pressure balance, Eq. (\ref{eq39}), we
obtain
\begin{equation}
\label{eq50}
\tilde{P}_{a}(1-Ma_{a}^{2})=\tilde{P}_{c}(1-Ma_{c}^{2}).
\end{equation}
Equation (\ref{eq50}) indicates the critical role of flow
compression in the extreme CJ deflagration regime with
$Ma_{c}^{2}=1$. In the CJ regime, sound in the hot plasma
propagates to the front as fast as it is drifted from the front by
the flow. As a result, pressure perturbations in the cold plasma
cannot be balanced by similar perturbations in the hot plasma. So,
one should expect that the CJ regime is a critical one for the DL
instability. For comparison, assuming extra condition (\ref{eq40})
like in \cite{Kadowaki-1995,Piriz-2001}, one obtains "$+$" instead
of "$-$" in both sides of Eq. (\ref{eq50}), which means nothing
special for the CJ deflagration. The special role of the CJ regime
has been also indicated in \cite{He-2000}.

\textbf{\it{C. Solution to the stability problem}}

\begin{figure}
\includegraphics[width=0.9\columnwidth]{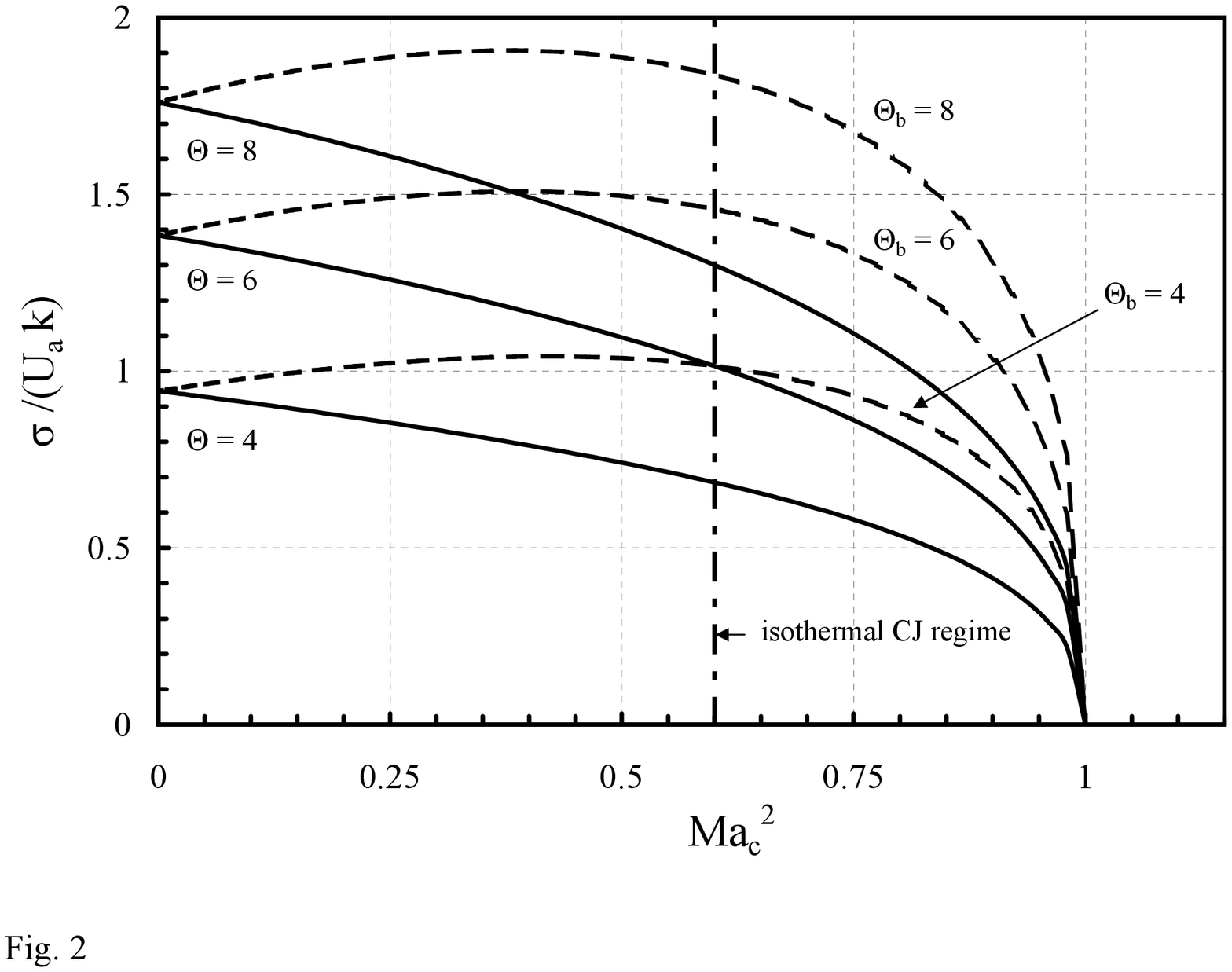}
\caption{Scaled instability growth rate $\Gamma = \sigma / U_{a}k$
versus the Mach number in hot plasma for the fixed expansion factors
$\Theta = 4, 6, 8$. The dashed lines show similar solutions for the
fixed heating factors $\Theta_{b} = 4, 6, 8$. The dashed-dotted line
indicates the isothermal CJ regime of laser ablation.} \label{fig-2}
\end{figure}

The modes in the uniform flows (\ref{eq31}), (\ref{eq32}),
(\ref{eq36}) and the matching conditions at the deflagration front
(\ref{eq37}), (\ref{eq39}), (\ref{eq49}) determine the solution to
the spectral problem for  $\Gamma \equiv \sigma / U_{a}k$. After
heavy but straightforward algebra presented in the Appendix one can reduce this set of
equations to
\begin{equation}
\label{eq51} (\Gamma \eta_{s} - \Theta)(\Gamma +  \eta_{a}) +
(\Gamma - \Theta \eta_{s})\left[1+
\eta_{a}\left(\frac{\Gamma}{\Theta}-\frac{\Theta -
1}{\Gamma}\right)\right]=0,
\end{equation}
where the following designations are introduced
\begin{equation}
\label{eq52} \eta_{a}= \sqrt{1+ Ma_{a}^{2}(\Gamma^{2}-1)} ,
\end{equation}
\begin{equation}
\label{eq53} \eta_{s}= \sqrt{1+
Ma_{c}^{2}\left(\frac{\Gamma^{2}}{\Theta^{2}}-1\right)} .
\end{equation}
In the limit of small Mach  number Eq. (\ref{eq51}) reduces to the classical DL result.
Numerical solution to (\ref{eq51}) is shown in Fig. \ref{fig-2}
 versus the Mach number for different values of the expansion
factor $\Theta = 4, 6, 8$  and the adiabatic exponent  $\gamma =
5/3$. For all expansion coefficients the instability growth rate
decreases monotonically with the Mach number and it turns to zero
in the adiabatic CJ regime of extremely fast deflagration at
$Ma_{c}^{2}=1$. Laser ablation in inertial confined fusion
corresponds to the isothermal CJ regime with
$Ma_{c}^{2}=1/\gamma$. In that case the DL instability is
non-zero, but it is considerably weaker than in the classical
incompressible case. This is different from the previous results
of Ref. \cite{Kadowaki-1995} predicting stronger DL instability at a
discontinuous combustion front in a compressible flow. This is
also different from the results of Ref. \cite{Piriz-Portugues-2003}
predicting no influence of compression effects for a discontinuous
front of laser deflagration/ablation. According to Ref. \cite{Piriz-Portugues-2003}, compression
effects come to play only as the second-order terms proportional
to the finite deflagration thickness.
On the contrary, we obtained decrease of the DL instability growth rate because of plasma
compression  already
for a discontinuous deflagration front. Taking into account finite
deflagration thickness we will, presumably, find strong
influence of compression effects too.

The numerical studies \cite{Travnikov.et.al-1997,Travnikov.et.al-1999} of the DL instability of a fast flame require a separate comment. The numerical data of Refs. \cite{Travnikov.et.al-1997,Travnikov.et.al-1999} demonstrated  increase of the DL instability because of the compression effects for a fast flame of finite thickness with the Arrhenius reaction of the first order. Comparing the present theory to the numerical results of Refs. \cite{Travnikov.et.al-1997,Travnikov.et.al-1999} one should notice that papers \cite{Travnikov.et.al-1997,Travnikov.et.al-1999} used the heating factor (temperature ratio) $\Theta_{b}=T_{c}/T_{a}$ as a fixed parameter instead of the expansion factor (density ratio) $\Theta=\rho_{a}/\rho_{c}$ used in the present paper. The heating factor is related to the expansion factor as
\begin{equation}
\label{eq-heating} \Theta_{b}= \Theta- \frac{ \gamma \Theta (\Theta -1)Ma_{c}^{2}}{\Theta + \gamma
(\Theta - 1) Ma_{c}^{2}}.
\end{equation}
The parameters $\Theta_{b}$ and $\Theta$ coincide for an incompressible flow, but they differ considerably when compression is strong. Choosing fixed $\Theta_{b}$ or $\Theta$ have quite different physical meaning, and we have to decide which parameter is more appropriate for the investigation. In a compressible flow, the deposited energy goes partly into the thermal energy (heating) and partly into the kinetic energy of the flow, which makes temperature ratio $\Theta_{b}$ smaller than the density ratio $\Theta$, see Eq. (\ref{eq-heating}). Still, the DL instability develops because of the density ratio: strong heating without expansion cannot produce the DL instability. Respectively, trying to keep the heating factor $\Theta_{b}$ fixed in the studies at large values of the Mach number, we inevitably have to increase the expansion factor $\Theta$. Larger densities ratio $\Theta$ produces stronger DL instability, which makes a deceitful  impression that compression effects enhance the DL instability. Such an approach may be, probably, justified for the combustion studies, but it definitely does not hold for laser ablation in inertial confined fusion.
In the case of laser ablation, the density ratio $\Theta$ in plasma is specified by the frequency of laser light, which makes the condition of fixed expansion factor $\Theta$ quite natural. For comparison, we have plotted solution to the spectral problem, Eq. (\ref{eq51}), for the fixed heating factor $\Theta_{b}$. The respective plots are presented in Fig. \ref{fig-2} by the dashed lines. The dashed lines do show some increase of the instability growth rate at small and moderate values of the Mach number similar to the numerical data of Refs. \cite{Travnikov.et.al-1997,Travnikov.et.al-1999}. The numerical simulations Refs. \cite{Travnikov.et.al-1997,Travnikov.et.al-1999} have been performed for combustion fronts of finite thickness. In order to compare the present case to Refs. \cite{Travnikov.et.al-1997,Travnikov.et.al-1999} quantitatively, we also have to take into account finite thickness of the ablation region in inertial confined fusion.
This is the subject for
future work, which is in progress now.

At the end of this subsection, we would like to point out that complete decrease of the DL instability in the CJ regime has an
interesting physical interpretation. It is well-known that a
deflagration front cannot propagate faster than the CJ
deflagration \cite{Landau-Lifshitz-Fluid}. On the other hand, it
is also well-known that the DL instability increases the
deflagration velocity at the nonlinear stage
\cite{Bychkov-Liberman-2000,Kadowaki-Hasegawa-2005}. Then, what
kind of nonlinear outcome we may expect for the DL instability at
the CJ deflagration? The only possibility is that there is no DL
instability in the CJ regime at all;   we have
obtained the same result solving the stability problem.

\textbf{\it{D. Influence of source localization}}

We have to make one more comment on the validity domain of the
present results. In Sec. II we considered only
localized energy sources  $\Omega_{R}$, which turn to zero in the
hot plasma together with the first derivatives. This restriction
is important for the present analysis. As a counter-example, let
us consider a hypothetic energy source in the form
\begin{equation}
\label{eq54} \Omega_{R}= \Omega_{a}(\rho - \rho_{c})^{\nu}\exp\left(-
\beta \frac{\rho}{\rho_{c}}\right).
\end{equation}
Here  $\rho_{c}$ is the final density of hot plasma, which is
similar to density at the critical surface, the factor  $\beta
>>1$ plays the same role as Zeldovich number (scaled activation
energy) in an Arrhenius reaction, $\nu$ imitates the reaction order
and $\Omega_{a}$  is some
constant. The function (\ref{eq54}) is constructed taking into
account similarities with the Arrhenius reaction. In the case of  the first-order reaction $\nu=1$ and large Zeldovich number $\beta
>>1$ the function
(\ref{eq54}) is localized, but the first derivative of
(\ref{eq54}) is not. Obviously, $\partial\Omega_{R}/\partial\rho$
is non-zero in the hot plasma with $\rho = \rho_{c}$ for $\nu=1$, which
modifies the perturbed equation (\ref{eq23}). As a consequence,
the structure of sound modes in the hot plasma is modified too. In
that case, within the discontinuity approach of $kL_{c}<<1$, the
leading term in (\ref{eq23}) is
$(\partial\Omega_{R}/\partial\rho)\tilde{\rho}=0$. This also leads
to $\tilde{\rho}_{c}=0$ in the hot plasma instead of the adiabatic
relation (\ref{eq25}). We also obtain $\mu_{c}=-k$ instead of
(\ref{eq29}); and the structure of the acoustic wave in the hot plasma
coincides with the incompressible case. The instability
growth rate in that case is almost the same as in the
incompressible limit: the difference is only in the small terms
$Ma_{a}^{2}<<1$. As we can see, taking the energy source in the
form (\ref{eq54}) with $\nu=1$ we come to the results close to Ref.
\cite{Piriz-Portugues-2003} and different from the main
conclusions of the present paper. Unfortunately, the work
\cite{Piriz-Portugues-2003} did not specify details of the energy
release in the analysis.
As another example, we can point out the
numerical solution Refs.
\cite{Travnikov.et.al-1997,Travnikov.et.al-1999}. The solution
of Ref. \cite{Travnikov.et.al-1997,Travnikov.et.al-1999} involved energy
release described by an additional differential equation
corresponding to a hypothetic first-order Arrhenius reaction.
Similar to (\ref{eq54}) with $\nu=1$, perturbations of the energy source in Refs.
\cite{Travnikov.et.al-1997,Travnikov.et.al-1999} were not
localized within the deflagration front. On the contrary, taking the second order $\nu=2$ in Eq. (\ref{eq54}) instead of the first one, we have the energy release strongly localized together with its derivatives. For $\nu > 1$ in Eq. (\ref{eq54}) we recover the main results of the present analysis.
Therefore, solution to the stability problem depends on the type
of energy release. If the energy release is localized, then we
obtain reduction of the DL instability by the compression effects
shown in Fig. \ref{fig-2}. If the source is not sufficiently
localized, then we may obtain increase or decrease of the
instability growth rate, or no effect at all, similar to the
previous works
\cite{Kadowaki-1995,Piriz-Portugues-2003,Travnikov.et.al-1997,Travnikov.et.al-1999}.
This reasoning is of special concern for the work \cite{He-2000},
where the additional condition was not formulated as an equation
at the front; but it took the form of an extra condition imposed
in the flow of hot gas (plasma). One cannot rule out, that the
assumption of zero entropy perturbations adopted in Ref. \cite{He-2000} holds for a certain type of
energy source.

\begin{figure}
\includegraphics[width=0.9\columnwidth]{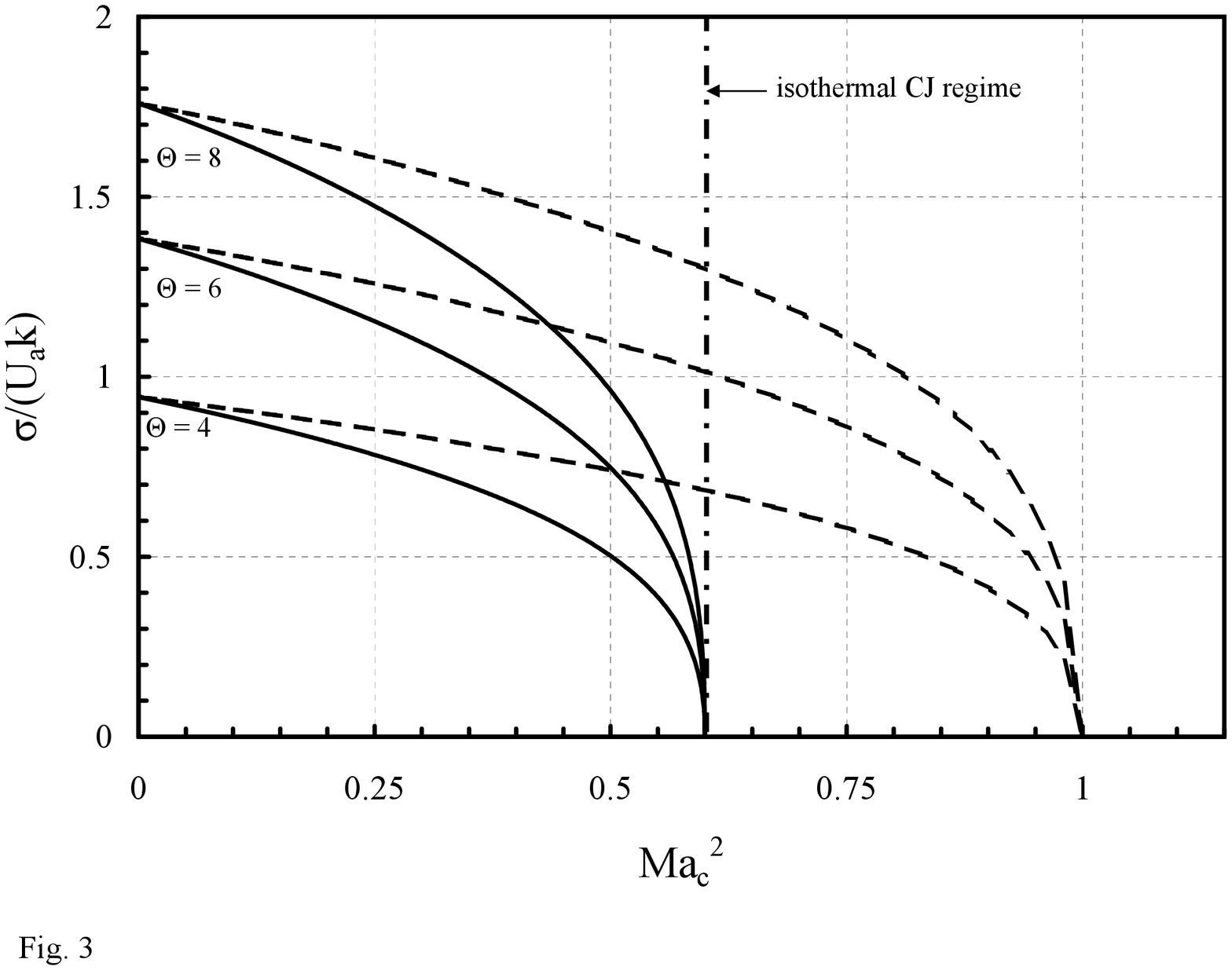}
\caption{Scaled instability growth rate $\Gamma = \sigma / U_{a}k$
versus the Mach number in hot plasma for the fixed expansion factors
$\Theta = 4, 6, 8$ in the case of adiabatic (dashed) and isothermal (solid) sound perturbations behind the front.  The dashed-dotted line indicates the
isothermal CJ regime of laser ablation.} \label{fig-3}
\end{figure}

A referee of the present work suggested also considering the case of isothermal sound in
the uniform flow of hot plasma, $\widetilde{T}_{c}=0$, which may be obtained by appropriate modifications of the energy source. Such a condition is in line with the idea of isothermal plasma corona for a planar ablation flow \cite{Manheimer.et.al-1982,Meyer-Thiell-1984,Fabbro.et.al-1985}. In that case Eq. (\ref{eq25}) should be replaced by $\widetilde{P}=(P/\rho)\widetilde{\rho}$ in the hot plasma and $Ma_{c}^{2}$ by $\gamma Ma_{c}^{2}$ in Eqs. (\ref{eq29}), (\ref{eq50}) and (\ref{eq53}). Respective solution to the stability problem is shown in Fig. \ref{fig-3}. In that case complete stabilization of the DL instability happens already in the isothermal CJ regime, which is the hydrodynamic regime of laser ablation. We stress that such a result is not general, but it holds only for a specific type of energy source. Any attempt to obtain the same result in general by appealing to strong thermal conduction in plasma corona is incorrect as long as one works within the model of a discontinuous front. We remind that the condition of strong thermal conduction for perturbations implies $k L_{c}>>1$, which is just the opposite to the condition of a discontinuous front $k L_{c}<<1$. Taking into account finite deflagration thickness and thermal conduction like in
 \cite{Searby-Clavin-1986}, we can evaluate the cut-off wavelength of the ablation DL instability to be an order of magnitude larger than $L_{c}$. Therefore, the DL instability develops on length scales much larger than the distance from the ablation surface to the critical one. At such large length scales one may not treat sound perturbations as isothermal ones.

Thus we come to the question if the energy source is
sufficiently localized in laser fusion. The spacial damping rate
of the laser light intensity is \cite{Eliezer}
\begin{equation}
\label{eq55} \frac{dI}{dz}=KI,
\end{equation}
with the absorption coefficient
\begin{equation}
\label{eq56} K\propto
\frac{n^{2}}{T^{3/2}}\left(1-\frac{n}{n_{c}}\right)^{-1},
\end{equation}
where  $n$ is electron concentration and $n_{c}$  is the critical
value, for which plasma frequency is equal to the laser frequency
\begin{equation}
\label{eq57} \omega^{2}=\omega_{p}^{2}= \frac{
e^{2}n_{c}}{m_{e}\varepsilon_{0}}.
\end{equation}
Mark that laser radiation in (\ref{eq55}) propagates in the
negative direction. Electron concentration in plasma is proportional to density  $n = \rho /M$,
where $M$ is mass per one electron. Localization of energy source
in (\ref{eq55}) is due to the singularity of the damping rate
(\ref{eq56}) at  $n = n_{c}$,  $\rho = \rho_{c}$. If we consider
derivatives of the damping rate
\begin{equation}
\label{eq58} \frac{\partial K}{\partial T} = -\frac{3K}{2T},
\end{equation}
\begin{equation}
\label{eq59} \frac{\partial K}{\partial n} = \frac{2K}{n}+
\frac{K}{2(n-n_{c})},
\end{equation}
we find localization as strong as in (\ref{eq56}) or even
stronger. Thus, energy source in laser ablation is indeed strongly
localized together with its derivatives, and the results of Fig.
\ref{fig-2} hold for laser ablation.


\section{IV. Summary}

In the present work we have solved the problem of the DL
instability at a fast deflagration in a compressible flow. The
solution is obtained within the traditional model of a
discontinuous front. Still, the approach of a discontinuous
deflagration front suffers from the deficit of boundary
conditions. We derive the missing condition as an eigenvalue of
the equation of energy transfer. The derived condition corresponds
to a constant mass flux at the deflagration front. In the limit of
incompressible flow it goes over to the classical DL condition. We
demonstrate that solution to the problem depends on the type of
energy source. In the common case of a strongly localized source,
compression effects make the DL instability considerably weaker.
In particular, the DL instability growth rate is reduced for laser
ablation in comparison with the classical incompressible case. The
instability disappears completely in the Chapman-Jouguet regime of
deflagration. However, if the energy source is not sufficiently
localized, then it may influence properties of the sound
perturbations in the hot plasma behind the front. In that case the
properties of the instability also depend on a particular type of
the energy source.

\section{Acknowledgements}

This work has been supported in part by the Swedish Research Council
(VR) and by the Kempe Foundation.

\renewcommand{\theequation}{A\arabic{equation}}
\setcounter{equation}{0}  
\section{Appendix: Derivation of Eq. (\ref{eq51}).}

Here we present short derivation of Eq. (\ref{eq51}). The spectral problem is determined by the equations (\ref{eq29}) - (\ref{eq31}), (\ref{eq36}) - (\ref{eq39}), (\ref{eq49}). Equation (\ref{eq39}) may be also replaced by  (\ref{eq50}), which is more concise.
Equations (\ref{eq37}), (\ref{eq49}) lead to
\begin{equation}
U_{a}\widetilde{P}_{a}/c_{sa}^{2}+\rho_{a}(\widetilde{u}_{za}-\Gamma
U_{a}kf)=0 \label{A2}
\end{equation}
\begin{equation}
U_{c}\widetilde{P_{c}}/c_{sc}^{2}+\rho_{c}(\widetilde{u}_{zc}-\Gamma
U_{a}kf)=0 \label{A3}
\end{equation}
We perform the following steps in calculations:

  1) $\widetilde{P}_{c}$ is expressed through $\widetilde{P}_{a}$
    and $\widetilde{u}_{za}$ using (\ref{eq30}), (\ref{eq50});

  2) $kf$ is coupled to $\widetilde{u}_{za}$ using
    (\ref{A2}), (\ref{eq30});

  3) using $kf$ and (\ref{A3}), we  relate $\widetilde{u}_{zc}$ and $\widetilde{u}_{za}$;

  4) using (\ref{eq38}), (\ref{eq31}), (\ref{eq30}), (\ref{eq49}), we relate $\widetilde{u}_{xc}$  and $\widetilde{u}_{za}$;

  5) everything is substituted into (\ref{eq36}).

Steps 1) - 4) in the calculations lead to
\begin{equation}
  \widetilde{P}_{c}=\alpha\widetilde{P}_{a}=-\alpha\beta
  \rho_{a}U_{a}\widetilde{u}_{za}, \label{A4}
\end{equation}
\begin{equation}
  \Gamma U_{a}kf=(1-Ma_{a}^{2}\beta)\widetilde{u}_{za},
  \label{A5}
\end{equation}
\begin{equation}
  \widetilde{u}_{zc}=(1-Ma_{a}^{2}\beta+\alpha\beta Ma_{c}^{2})\widetilde{u}_{za},
  \label{A6}
\end{equation}
\begin{equation}
  \widetilde{u}_{xc}=i\widetilde{u}_{za}\left(\frac{k}{\mu_{a}}-
    \frac{\Theta-1}{\Gamma}(1-Ma_{a}^{2}\beta)\right),
  \label{A7}
\end{equation}
where the following designations have been introduced for $\alpha$ and $\beta$:
\begin{eqnarray}
 \alpha=\frac{1-Ma_{a}^{2}}{1-Ma_{c}^{2}},\; \;\;\;\;
 \beta=\frac{\Gamma k}{\mu_{a}}+1.
\end{eqnarray}
Substituting (\ref{A4})-(\ref{A7}) into
Eq. (\ref{eq36}) we obtain solution to the spectral problem in the form of a single algebraic equation
\begin{equation}
  -\alpha\beta\left(\frac{k}{\mu_{c}}+\frac{\Gamma}{\Theta}\right)
  \left(\frac{k\Gamma}{\Theta\mu_{c}}+1\right)^{-1}+
  \frac{\Gamma}{\Theta}\left(1-Ma_{a}^{2}\beta+\alpha\beta Ma_{c}^{2}\right)+
  \left(\frac{k}{\mu_{a}}-\frac{\Theta-1}{\Gamma}(1-Ma_{a}^{2}\beta)\right)=0.
  \label{A8}
\end{equation}
This equation may be further simplified. We introduce auxiliary designations:
\begin{equation}
  \frac{\mu_{a}}{k}=\Gamma\frac{Ma_{a}^{2}}{1-Ma_{a}^{2}}
   +\frac{\eta_{a}}{1-Ma_{a}^{2}},\;\;\;\;
  \eta_{a}=\sqrt{1+Ma_{a}^{2}(\Gamma^{2}-1)},
  \label{A9}
\end{equation}
\begin{equation}
  \frac{\mu_{c}}{k}=\frac{\Gamma}{\Theta}\frac{Ma_{c}^{2}}{1-Ma_{c}^{2}}
   -\frac{\eta_{c}}{1-Ma_{c}^{2}},\;\;\;\;
  \eta_{c}=\sqrt{1+Ma_{c}^{2}(\Gamma^{2}/\Theta^{2}-1)},
  \label{A10}
\end{equation}
which leads to
\begin{equation}
  \left(\frac{k}{\mu_{c}}+\frac{\Gamma}{\Theta}\right)
  \left(\frac{k\Gamma}{\Theta\mu_{c}}+1\right)^{-1}\!\!\!=
  \frac{\Gamma\mu_{c}+\Theta}{\Gamma+\mu_{c}\Theta}=
  \frac{\Gamma^{2}Ma_{c}^{2}-\Theta\Gamma\eta_{c}+\Theta^{2}(1-Ma_{c}^{2})}
  {\Theta(\Gamma-\Theta\eta_{c})}. \label{A11}
\end{equation}
Substituting (\ref{A9})-(\ref{A11}) into (\ref{A8}) and
multiplying  by $(\Gamma-\Theta\eta_{c})$ we obtain
\begin{equation}
  \alpha\beta(\Gamma\eta_{c}-\Theta)(1-Ma_{c}^{2})+(\Gamma-\Theta\eta_{c})
  \left[\frac{k}{\mu_{a}}+(1-Ma_{a}^{2}\beta)
  \left(\frac{\Gamma}{\Theta}+\frac{\Theta-1}{\Gamma}\right)\right]=0.
  \label{A12}
\end{equation}
Taking into account  that
\begin{eqnarray}
  \beta=\frac{k}{\mu_{a}}\frac{\Gamma+\eta_{a}}{1-Ma_{a}^{2}},\;\;\;
  \;\;\;(1-Ma_{a}^{2}\beta)=\frac{\eta_{a}k}{\mu_{a}},
\end{eqnarray}
we come to the final equation (\ref{eq51})
\begin{eqnarray}
  (\Gamma\eta_{s}-\Theta)(\Gamma+\eta_{a})+(\Gamma-\Theta\eta_{s})\left[1+\eta_{a}
  \left(\frac{\Gamma}{\Theta}-\frac{\Theta-1}{\Gamma}\right)\right]=0.
\end{eqnarray}
This gives solution to the spectral problem in a form of one concise algebraic equation.



\end{document}